\documentclass[sigconf, nonacm]{acmart}
\pdfoutput=1
\makeatother
\usepackage{booktabs}
\usepackage{hyperref}
\usepackage{mdframed}

\usepackage{flushend}
\usepackage{natbib}

\usepackage{amsmath}
\usepackage{amsfonts}
\usepackage{algorithmic}
\usepackage{xcolor}
\usepackage{xspace}
\usepackage{tabularx}

\usepackage{makecell}
\usepackage{booktabs}

\hypersetup{
  colorlinks,
  citecolor=green,
  linkcolor=red,
  urlcolor=blue}
\usepackage{listings}
\usepackage[draft=true]{minted}
\usepackage{etoolbox}
\makeatletter
\AtBeginEnvironment{minted}{\dontdofcolorbox}
\def\dontdofcolorbox{\renewcommand\fcolorbox[4][]{##4}}
\makeatother

\usepackage{graphicx}
\usepackage{listings}
\usepackage{url}

\newcommand{\ttt}[1]{\mintinline[fontsize=\small]{bash}{#1}}

\newmdtheoremenv[%
  outerlinewidth=2pt,
  leftmargin=0em,
  rightmargin=0em,
  innertopmargin=0pt,
  splittopskip=\topskip,
  skipbelow=\baselineskip,
  skipabove=.5\baselineskip]{implication}{Implication}

\begin{document}

%
\title{Quieting the Static:\\
A Study of Static Analysis Alert Suppressions}


\author{Georgios Liargkovas}
\orcid{0000-0002-8896-4044}
\affiliation{%
  \institution{AUEB, Greece}
  \country{}
}
\author{Evangelia Panourgia}
\affiliation{%
  \institution{AUEB, Greece}
  \country{}
}
\author{Diomidis Spinellis}
\orcid{0000-0003-4231-1897}
\affiliation{%
  \institution{AUEB, Greece \& }
  \country{TU Delft, Netherlands}
}
\begin{abstract}
Static analysis tools are commonly used to detect defects
before the code is released.
Previous research has focused on their overall effectiveness
and their ability to detect defects.
However, little is known about the usage patterns of warning suppressions:
the configurations developers set up in order to prevent the appearance
of specific warnings.
We address this gap by analyzing
how often are warning suppression features used,
which warning suppression features are used and for what purpose,
and also
how could the use warning suppression annotations be avoided.
To answer these questions
we examine 1\,425 open-source Java-based projects that utilize
Findbugs or Spotbugs for warning suppressing configurations and source
code annotations.
We find that although most warnings are suppressed,
only a small portion of them get frequently suppressed.
Contrary to expectations,
false positives account for a minor proportion of suppressions.
A significant number of suppressions introduce technical debt,
suggesting potential disregard for code quality
or a lack of appropriate guidance from the tool.
Misleading suggestions and incorrect assumptions also lead to suppressions.
Findings underscore the need 
for better communication and education 
related to the use of static analysis tools,
improved bug pattern definitions,
and better code annotation.
Future research can extend these findings 
to other static analysis tools, 
and apply them to improve the effectiveness of static analysis.
\end{abstract}



\maketitle

\section{Introduction}

Static analysis is a popular and powerful weapon
that can be used to identify many types of software bugs
without running the actual code~\cite{ayewah2008using}. 
In contrast to dynamic analysis, static analysis can identify bugs 
in earlier development stages, 
before they emerge as expensive~\cite{7476667} run-time errors,
while usually requiring less computational power 
(depending on the analysis algorithm)~\cite{babati2017static}.
For this reason, it has been used extensively 
as an extra layer of protection
by the majority of software developers 
in most open-source and proprietary software projects~\cite{7476667}.

Despite the above, static analysis can be both a blessing and a curse.
Applying even the most advanced static analyzer
does not ensure that all possible defects have been discovered~\cite{babati2017static}.
Furthermore, false positive 
warnings remain a common issue among static analysis tools~\cite{chess2004static},
notifying developers about inexistent errors.
Such cases may lead to unnecessary development efforts 
and development slow-downs,
and have been identified as one of the key reasons 
many software developers 
avoid using static analysis~\cite{johnson2013don}.

As a means of preventing static analysis warnings 
from overwhelming software practitioners,
most modern static analysis tools have
introduced ways of dealing with false positives
by modifying the sensitivity of diverse matchers, 
or even completely ignoring the respective output, 
through configuration files and special annotations~\cite{ayewah2008using}.
Throughout this work we refer to these mechanisms as \textbf{warning suppression features}.
These features can be used to silence or 
even completely skip the analysis in some parts of the code,
indicating that the developer willingly ignores any warnings
that could possibly occur inside the suppressed code, 
deeming the warning related to that code \textbf{unactionable} 
(alerts that do not require developer changes)~\cite{heckman2007adaptively}.

While substantial previous work
has focused on the identification, characteristics and frequency 
of false positive alerts~\cite{park2016battles,8622456,podelski2016classifying}, 
the discovery of ignored static analysis patterns~\cite{hanam2014finding,reynolds2017identifying}, 
and their empirical examination,
warning suppression is a feature that is still being overlooked 
by static analysis evaluation studies.
This paper extends existing work 
by extracting insights about
unactionable static analysis alerts 
through various warning suppression features,
specified in open source software projects.
Throughout our study, we employ Findbugs and Spotbugs,
a family of widely-used, feature-rich Java source code static analyzers,
that are able to detect more than 400 bug patterns across 12 distinct categories.

Not all bug patterns are created equal~\cite{lavazza2020empirical}.
From a look inside the annotations and configuration files 
of certain large open source projects
it is apparent that certain alert codes are suppressed more than others, 
indicating a possibly higher false positive ratio.

Based on these issues, this study is guided by the following research questions. 
We explore how often are warning suppression features used in static analysis (\textbf{RQ1}) but also 
which warning suppression features are used (\textbf{RQ2}).
However, the mere presence of suppression features 
in a project's configuration does not always imply 
that the suppressed warning is always unactionable or a false positive. 
This uncertainty makes us inquire for what purposes are static analysis warning suppression features employed (\textbf{RQ3}).
Finally, we aim to identify practices 
that could replace the use of warning suppression features, 
aiming to contribute to code quality improvements,
by asking how the use of static analysis warning suppression annotations 
could be avoided (\textbf{RQ4}).

This paper makes several contributions to 
the state of the art.
First, we offer a unique dataset
of Findbugs and Spotbugs warning suppression instances, 
for warning suppression configurations and annotations 
with 8\,298 and 2\,943 entries respectively, 
a resource that can be utilized in future research 
to enhance our understanding of warning suppression patterns. 
Second, we provide a quantitative examination 
of the use of warning suppressions, 
investigating which annotations are used 
and how frequently they are applied. 
This enables a deeper view of how and when 
developers use these tools. 
Third, we offer a qualitative exploration 
of the reasons developers utilize static analysis warning suppressions. 
This provides insights in identifying 
the motives behind their choices, 
promoting better tool design and developer education. 

\section{Background and Related Work} 
\label{subsec:backgroundandrelatedwork}

Findbugs and Spotbugs, 
key static analysis tools, 
are used in major projects, 
to detect error-prone code fragments. 
Despite their usefulness, 
they produce numerous false positives, 
prompting the use of warning suppression features.
Furthermore, extensive research 
has tackled false positives, 
with innovative models proposed to improve accuracy, 
such as analyzing the count of lines 
previously inserted 
and code-to-comments ratio.
Related work has also identified 
various causes 
of false positives
like W3C APIs, 
browser-specific APIs, 
and dynamic file loading. 
Conversely, our study uniquely uses 
warning suppression features 
to differentiate actionable 
from unactionable warnings.

\subsection{Static analysis tools}
Static analysis tools can be used 
to analyze a project's source code either on demand 
or as part of a project's continuous integration pipeline.
Furthermore, these tools are included 
in most modern integrated programming environments (IDEs),
providing real-time error detection.
The main advantages of static analysis lie in the fact that
it can be applied relatively early in the development lifecycle
while maintaining a relatively low performance cost 
compared to other code analysis methods
that operate at runtime~\cite{chess2004static}.
Static analysis is extensively used 
by large open source as well as proprietary software projects.
For example, Google has integrated static analysis tools 
into its software development pipeline, 
utilizing an array of style checkers (Checkstyle, Pylint, and Golint) 
and bug-finding tools (Error Prone, ClangTidy, Govet, and Checker Framework)~\cite{ayewah2008using,10.1145/1831708.1831738}.
Additionally, the majority of software developers in industry 
incorporate static analysis in their CI builds~\cite{8330195}.

However, although most modern static analysis tools
are generally sound, 
they also produce a substantial amount of false positives.
These include alerts that are too trivial to fix (e.g. indentation), 
unlikely runtime errors or completely incorrect types of warnings~\cite{liu2018mining}.
Various studies estimate that false positives account for 18\% to 86\% 
of total warnings depending on the tool used~\cite{kremenek2003z,shen2011efindbugs,lenarduzzi2021critical}.
False positives can become a significant burden to software developers,
especially in larger software projects~\cite{10.1145/1831708.1831738}.
Distinction between false and true positives is known 
to be a time consuming and error prone process~\cite{wedyan2009effectiveness}.
Most tools have introduced various ways of dealing with false positive warnings,
with the most popular solution being alert suppression, 
which however also requires significant human effort to configure correctly.
As a result, many organizations and software developers 
tend to avoid using static analysis in their projects 
because of the highly misleading unactionable warnings~\cite{6606613}.

\subsection{Findbugs and Spotbugs}

Findbugs\footnote{\url{https://findbugs.sourceforge.net/}} is an open-source 
bug finding tool for Java programs.
It utilizes a set of ad-hoc methods to identify
various error-prone code fragments~\cite{ayewah2008using}.
Findbugs utilizes a combination of syntactic and 
dataflow analysis, 
applied directly to a Java program's bytecode, 
to detect defectful patterns.
It allows the detection of a wide range of issue types, 
including \emph{bad practice}, 
\emph{correctness},
\emph{multithreaded correctness}
\emph{style}, 
\emph{performance},
\emph{malicious code},
\emph{internationalization} and
\emph{security}.
Moreover, it allows the user to configure various aspects of the analysis
(i.e. category of analysis, confidence, etc.)
conducted, 
through command-line arguments, 
configuration files,
and source code annotations.

The functionality of Findbugs can be easily expanded
through various plug-ins written by third parties.
The tool has been around for more than a decade,
establishing itself as one of the most popular 
open-source tools used for static analysis in Java.
Findbugs is one of the most studied static analysis tools 
in the literature~\cite{kim2007warnings,gosain2015static,habib2018many,6606613,KWL22}.

Spotbugs\footnote{\url{https://spotbugs.github.io/}} 
is a popular Findbugs fork, 
offering support for the latter,
which has now been abandoned. 
Spotbugs extends Findbugs's functionality,
being able to detect more than 400 types of defects by default.

\subsection{Evolution of Static Analysis}
\label{subsec:related_work}

Static analysis tools have been evaluated 
in terms of the quality and taxonomy of their results. 

The study of false positive issues 
in static analysis tools has seen significant work 
over the past decades.
Kremenek and Engler~\cite{kremenek2003z} 
marked the inception of this research trajectory, 
where they employed the z-ranking technique 
to evaluate the accuracy of error reports 
produced by automated static analysis tools.
Building on this seminal work, 
advancements were made 
by Ayewah et al. \cite{ayewah2007evaluating} 
and Heckman et al. \cite{heckman2007adaptively}. 
Imtiaz and Williams \cite{10.1145/3314058.3317295} 
further contributed to the body of knowledge 
by analyzing the rate of warnings and false positives 
generated by Coverity Scan. 
Their study dissected Coverity analysis reports of prominent projects, 
pinpointing the most frequent triggers of unactionable warnings.
Ayewah et al. expanded the scope of prior work 
to examine various categories of Findbugs warnings 
within industry codebases such as Google and Sun, 
categorizing the alerts into three groups: 
\emph{false positives}, 
\emph{small issues}, and 
\emph{serious issues}. 
Concurrently, Heckman proposed an innovative 
adaptive model for rectifying inaccuracies 
in static analysis reports, 
considering variables such as 
user annotation suppressions and historical data~\cite{heckman2007adaptively}.
She later proposed the count of lines previously inserted in a file
as a means of predicting the actionability 
of static analysis warnings~\cite{10.1109/ICST.2009.45}.
Similarly, Liang et al. proposed code-to-comments ratio
for the same purpose~\cite{10.1145/1858996.1859013}.
Based on this prior work Wang et al.~\cite{10.1145/3239235.3239523}
conducted a systematic literature review
through which they identified a set of 23 features
that contribute to the actionability of
static analysis warnings.
Most recently, Kang et al.~\cite{KWL22} 
conducted an evaluation 
of these static analysis false-positive criteria
and discovered that the performance of most
of these metrics
was due to data leakage 
and data duplication issues,
with their true accuracy being lower 
than initially presumed.
The authors also outline the need of an 
unactionable warning identification suite.
Our study takes a different approach, 
focusing exclusively on warning suppressions,
a feature not examined in prior work, 
to draw conclusions about 
actionable and unactionable 
false positive static analysis warnings 
on a large scale.

Additionally, previous work has tried 
categorizing the identified false positive warnings.
Park et al.~\cite{park2016battles} 
evaluated 30 JavaScript web-based applications 
using the SAFE static analyzer.
They identified seven main causes 
of false positive warnings that they then reduced to 
four main categories: 
W3C APIs, 
browser-specific APIs, 
JavaScript library APIs, and 
dynamic file loading.
Similarly, Reynolds et al.~\cite{reynolds2017identifying} 
collected, evaluated, and documented 
false positive messages
produced by three automated static analysis tools 
on C and C++ code.
They composed a hierarchical list of causes 
that false positives can be reduced into, 
including memory allocation, 
array-related issues, 
buffer related issues, 
and predictable conditionals, 
among others.
Furthermore, de Mendonça et. al~\cite{de2013static} 
systematically mapped 64 static analysis result evaluation papers,
concluding that the majority of them 
(34\%) categorize warnings between true and false positives,
while 20\% of them provided 
a more in-depth explanation 
about the causes of the latter ones. 

\section{Methods}
\label{sec:methods}

\begin{figure}[t]    
    \includegraphics[width=\linewidth]{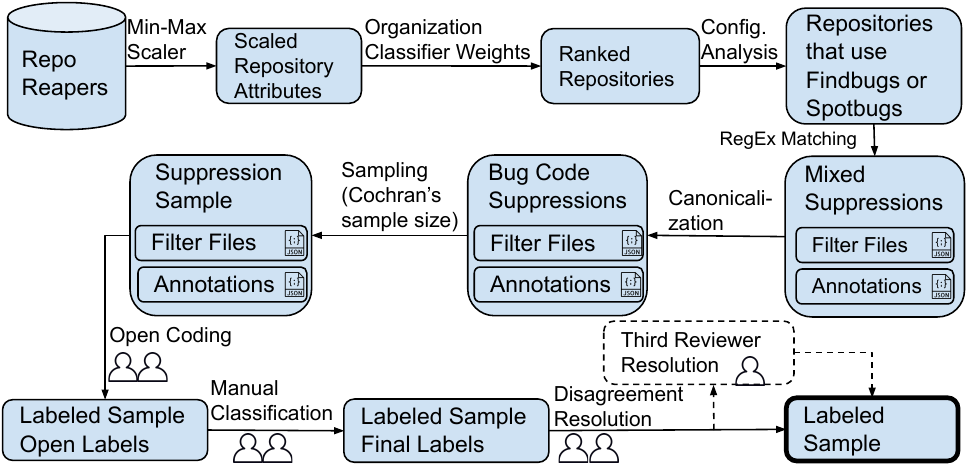}
    \caption{Research method overview.}
    \label{fig:methodology}
\end{figure}

Our workflow, depicted in Figure~\ref{fig:methodology}, 
primarily involves the selection
of static analysis tools, 
project selection, 
repository analysis, 
match canonicalization, 
and open coding. 
Findbugs and Spotbugs 
were chosen based on their performance and popularity. 
For project selection, 
a dataset of engineered GitHub software projects 
was used to select repositories using Java 
and either of the two aforementioned tools.
Metrics of the RepoReapers dataset were scaled 
and then ranked based on the \emph{Organization} classifier.
The repositories were then analyzed for warning suppressing 
annotations and configurations. 
Selected repositories were scanned 
for Findbugs and Spotbugs dependencies 
until a threshold of 100 consecutive misses
was reached.
Regarding warning suppression identification, 
regular expressions were used
for speed and simplicity. 
Matches were canonicalized to reflect bug patterns.
Samples of 368 warning suppression configurations and 
340 annotations were manually analyzed.
Finally, to understand the rationale 
behind warning message suppression, 
manual coding was applied to two samples, 
resulting from matches found previously. 
This involved the examination of each bug pattern entry 
and corresponding source code, 
leading to the identification 
of several categories based 
on their application goal and potential avoidance strategies.

\subsection{Static analysis tool selection}
We selected Findbugs and its successor Spotbugs based on 
the tools’ performance and popularity.
Regarding performance,
from a usability perspective, Nachtigall et. al~\cite{nachtigall2019explaining} 
evaluated 14 static analysis tools using six criteria: 
understandable warning messages, 
fix support, 
false positives, 
user feedback,
workflow integration 
and specialized user interface.
They further conducted a systematic evaluation in a set of 
46 non-proprietary static analysis tools~\cite{nachtigall2022large}
using 36 criteria, including 
warning message quality, 
false positive occurrence and handling,
workflow integration and user interface.
They discovered that only a fraction of the available tools (14/46)
offer a concrete way of dealing with false positives (e.g. via configuration and annotations).
Findbugs and Spotbugs were found to satisfy the 
warning message, 
false positive, 
workflow integration 
and user interface criteria.
Similarly, Rutar at al.~\cite{1383122} 
compared five popular static analysis tools for Java, 
reaching the same conclusion.
They also found out that Findbugs had the broadest bug type coverage, 
while remaining relatively fast
compared to tools of similar scope, such as PMD.

From a popularity perspective,
Findbugs and Spotbugs 
have been used by many large software organizations,
such as Google~\cite{10.1145/1831708.1831738}, 
the Apache Software Foundation, 
and the Eclipse Foundation~\cite{marcilio2020spongebugs}.
Additionally, Findbugs and Spotbugs are considered 
some of the most used and most mature analyzers~\cite{lavazza2020empirical,gosain2015static,6606613}

The above led to the selection of Findbugs and Spotbugs
for this study.

\subsection{Project Selection}
\label{subsec:proj_selection}

We needed to analyze the configuration of Findbugs and Spotbugs found in real-world examples
written in Java.
For this, the RepoReapers dataset of engineered GitHub software projects was utilized.
The authors use the term \emph{engineered} to refer to repositories 
that follow patterns similar to these of large organizations,
and are usually user rather than developer-centric.
The dataset consists of a set of 1\,853\,197 deduplicated GitHub repositories,
evaluated in diverse software engineering practices, 
called dimensions~\cite{munaiah2017curating}.

To facilitate analysis, repositories were filtered and ordered using the following process.
The repository dimension metrics included in the dataset were scaled 
using the MinMaxScaler of the Scikit-learn library,\footnote{\url{https://scikit-learn.org}}
and ranked according to the proposed dimension weights 
used by RepoReapers's \emph{organization} classifier.
The derived score of each repository,
indicates the extent to which 
it follows the development principles of larger organizations.
Only Java-based repositories were selected 
as this matched what the two selected tools can process.

\subsubsection{Repository Identification}

Using the process described above, 
a set of 462\,173 ranked candidate repositories was identified.
Not all of these projects, however, remain accessible, 
as they might have been deleted
or made private by their administrators.
Such cases were ignored.
Additionally, it is not known beforehand whether Findbugs or Spotbugs 
are a part of each project's analysis lifecycle, 
therefore all candidate repositories were cloned and searched.

First, GitHub repositories were locally cloned 
using the \ttt{--depth 1} parameter 
to conserve space.
The configuration files of three major Java project build tools 
(\ttt{pom.xml} for Maven, 
\ttt{build.gradle} for Gradle, 
\ttt{build.xml} for Ant) were then scanned for Findbugs and Spotbugs,
using regular expressions. 
Repositories not using these tools were excluded. 
The process above was conducted in an online fashion,
working through the repositories ordered by score,
until a threshold of 100 continuous non-candidate repositories was reached.
We set this threshold based on time, space, and bandwidth constraints. 
This resulted in the cloning and search of 12\,720 repositories, 
yielding 1\,425 referencing Findbugs and Spotbugs.

\subsection{Repository Analysis}
The selected tools offer two different warning suppression features: 
warning suppression code-embedded annotations and 
warning suppression configurations (realized through exclude filter files). 
Both methods aim for the same outcome with varying mechanisms. 
Warning suppression annotations code-embedded location-specific 
rule-setting for variables, 
code blocks, methods, and classes. 
Warning suppression configurations are XML documents 
defining rules expressed independently of the code.
They also support logic operations and pattern matching 
for dealing with false positive patterns.

\subsubsection{Configuration File Analysis}
\label{subsubsec:filter_file_analysis}
Configurations can include or exclude specific bug codes. 
We focused on the latter. 
To isolate such files, 
we identified all exclude filter filenames 
specified in project build files 
and used these to find XML warning suppressing 
configurations in the repositories. 
Only XML files with the root tag \ttt{<FindbugsFilter>} were kept. 
Within theese, bug code filters are represented with the \ttt{<Match>} tag,
which contains parameters that are used to specify
match information such as the \emph{class}, \emph{method} or \emph{field} the filter is applied on,
as well as the specific bug \emph{pattern}, \emph{code}, or \emph{category} of the match.
More complex logical constructs are allowed (i.e.\emph{and}, \emph{or}),
although rarely encountered in practice. 
We parsed distinct cases of the \ttt{<Match>} tag, 
that mentioned at least one specific 
\emph{pattern}, \emph{code}, or \emph{category},
saving the output in a global JSON file.

\subsubsection{Annotation Analysis}
The information in annotations 
is similar to warning suppressing configurations, 
therefore, the method was alike.
Our initial approach was to use the abstract syntax tree 
of Java programs
to spot any annotations. 
This would guarantee correct identification of the annotation scope
(i.e. field, method, class, etc.).
However, in practice, this computationally heavy process
required the analysis
of the entire codebase to function properly.
Given the number of repositories analyzed, it was much simpler, 
and almost equally accurate to use 
a much faster approach based on regular expressions.
A regular expression matcher 
applied to all the candidate repositories' 
Java source code files (*.java) 
was used to identify warning suppression annotations. 
To include annotation scope information, 
the lines following the annotation declaration were parsed.
We identified classes, methods, and fields as major categories. 
Given the similarity between 
warning suppressing features (annotations and configurations), 
the former were stored into a similar JSON file.

\subsection{Match Canonicalization}
\label{subsec:match-canonicalization}

The warning suppression configuration and annotation analysis
identified 5\,139 and 2\,329 bug matching clauses respectively. 
These matches may not represent a unique bug pattern, 
given that the tools under study organize warnings 
into a hierarchical structure of increasing abstraction: 
bug patterns, codes, and categories. 
Each match could reference a set of these entities, 
which could in turn correspond to multiple bug patterns. 
To maintain consistency, 
matches were canonicalized 
to represent single 
or multiple bug patterns —- the most granular entity of a warning.

The tools' bug pattern structure 
was extracted using a Python script that scraped 
the manual page of Spotbugs and two popular 
Findbugs-and-Spotbugs-compatible plugins, 
\emph{find security bugs}\footnote{\url{https://find-sec-bugs.github.io/}}
which consists of 141 security vulnerability pattern identifiers,
and FB contrib.\footnote{\url{https://fb-contrib.sourceforge.net/}},
which contains a wide range of general bug pattern identifiers.
We placed the bug patterns of the aforementioned plugins into separate bug categories.
This procedure revealed 911 bug patterns 
belonging to 434 bug codes 
within 12 categories.

After gathering the required information,
matches previously found were converted 
into single bug pattern matches. 
Consequently, each match entry represented a single bug pattern, 
yielding 8\,298 bug patterns across 168 repositories in 
warning suppression configuration files, 
and 2\,943 in warning suppression annotations 
across 91 distinct repositories.

\subsection{Canonicalized Dataset}
\label{subsec:canonicalized-dataset_exclude}
\begin{figure}
  \begin{minted}[fontsize=\footnotesize, escapeinside=||,linenos,xleftmargin=17pt]{json}
  {
    "Id": 0,
    "Id_norm": 5,
    "Range": "Class",
    "Class": {
      "name": "com.github.susom.database.OptionsDefault"
    },
    "File": "findbugs-exclude.xml",
    "Repository": "susom/database",
    "Type": "Filter",
    "Bug": "DMI_DOH",
    "Code": "DMI",
    "Category": "CORRECTNESS",
  }
  \end{minted}
    \caption{Example of a canonicalized exclude filter JSON match entry.}
    \label{fig:canonicalized_json_exclude}  
  \end{figure}

Following project selection and repository analysis, 
a JSON-formatted dataset, 
including warning suppression configuration 
and annotation suppressions, 
was created (see Section~\ref{sec:availability}). 
The warning suppressing configurations
consisted of 8\,298 instances, 
each containing specific code exclusion attributes, 
as shown in Figure~\ref{fig:canonicalized_json_exclude}. 
Similarly, the 2\,943 warning suppressing annotations 
also contained these attributes, 
with the addition of a \emph{justification} field 
and the observed \emph{line number}. 
The two elements of the dataset allowed 
a view of both warning suppression configuration and annotation practices 
in the analyzed repositories, 
providing a basis for the subsequent steps 
of sampling and open coding,
while facilitating reuse and further analysis.

\subsection{Sampling and Open Coding}
\label{subsec:sampling-and-open-coding}
To investigate the reasons developers use 
warning suppression features
we applied manual coding~\cite{corbin1990grounded} 
to two samples, 
derived from the matches identified
in~Section~\ref{subsec:match-canonicalization}. 
We used Cochran’s 
sample size and correction formula~\cite{cochran1977sampling},
which yields a representative sample for proportions
of large populations:

$$ n_0 = \frac{z^2p(1-p)}{e^2}$$

where $n_0$ is the required sample size, 
$p$ the proportion of population (0.5—maximum
variability), 
$z$ is found in the $Z$ score table, 
and $e$ is the desired margin
of error (5\%).
For the warning suppression configurations (8\,298 bug patterns), 
the sample size was 368, 
while for warning suppression annotations (2\,943 bug patterns), 
it was 340.

Each bug pattern was independently examined and openly labeled 
by two of this paper’s authors.
This examination involved analysis of both the bug pattern itself 
and the corresponding source code within its scope.
The categorization of each pattern was undertaken 
in relation to its application rationale and possible avoidance strategies. 
Subsequent discussions helped refine the
categories identified by the authors into conceptually similar codes.

\begin{figure}[htbp]
  \includegraphics[width=\linewidth]{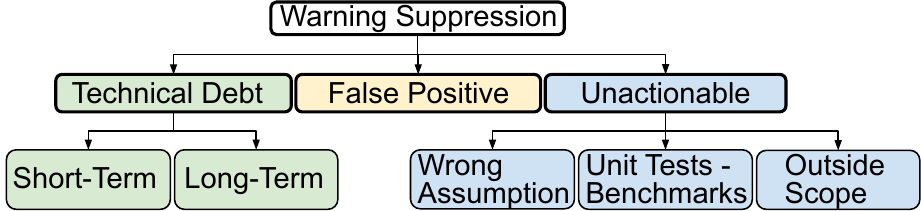}
  \caption{The hierarchical structure of the categories identified
  in the coding process.}
  \label{fig:category_structure}
\end{figure}

This process yielded three major categories and six subcategories 
(see Figure~\ref{fig:category_structure}). 
\emph{False positives} refer to incorrect issue identification by the static analysis tool.
The tool's error required these false alarms
to be suppressed in the codebase.
\emph{Technical debt} describes warnings that were correctly identified 
but ignored. 
This could be due to various reasons 
such as time constraints, 
budget issues, 
or even a lack of understanding of the problem. 
This category was subdivided into 
\emph{short-term technical debt} (easy-to-fix warnings) and 
\emph{long-term technical debt} (warnings requiring significant refactoring). 
\emph{Unactionable} instances include cases where the bug pattern matches the bug description, 
but there's no need for changes to the annotated part. 
This category was further divided into three subcategories: 
\emph{wrong tool assumptions}, 
\emph{testing}, 
and \emph{outside of analysis scope}.
\emph{Wrong tool assumptions} refer to cases where the match 
was based on incorrect assumptions made by the tool
but could not be described as false positives. 
\emph{Testing} represents scenarios 
related to testing or benchmarking, 
where dealing with most static analysis warnings is not easily justified.
\emph{Outside of analysis scope} cases include instances 
where the issue was not relevant to the current scope of analysis,
like automatically-generated files, resource files, etc.

A second round of categorization was conducted using these categories. 
Any disagreements were discussed, and if necessary (in very few cases), 
a third author with three decades of software development experience was consulted. 

To assess the reliability of the labeling between the two authors,
a measure of inter-rater agreement,
the Cohen's kappa ($\kappa$) coefficient~\cite{cohen1960coefficient}, was utilized. 
Given the six subcategories were equally likely to be assigned, 
the probability of random agreement $p_e$ 
was calculated as approximately $0.167$.
For the annotations, there was agreement 
on 310 out of 340 cases~(91\%), 
yielding an observed agreement $p_o$
of approximately $0.912$
and a $\kappa$ value of approximately $0.895$.
For the warning suppression configurations
agreement was reached on 311 out of 368 cases~(84\%), 
resulting in an observed agreement $p_o$ 
of approximately $0.842$. 
The $\kappa$ value for this category was found to 
be approximately $0.811$.
These values suggest 
a strong level of agreement between the two authors
during the initial categorization stage,
in both cases, 
indicating the reliability of the procedure. 

\section{Results}
\label{sec:results}

In brief, we found that warning suppression features are used in a small
subset of projects. The majority of them suppressed only a small number of times
although some outliers of commonly suppressed bug codes were found.
Most commonly suppressed warnings were related to broad categories
such as \emph{bad practice} and \emph{style} and \emph{correctness}.
Plugin-related warnings were also often suppressed
through annotations.
Additionally, contrary to our initial expectations,
exclusions of specific bug codes were more
suppressed at class scope, rather than at project level.
Finally, we found that the majority of the warnings
in both warning suppression features
were \emph{unactionable} and based on
\emph{wrong tool assumptions},
with \emph{outside of analysis scope} and \emph{testing}
being encountered less often.
Suppressions related to \emph{technical debt},
were also common, while \emph{false positives} 
attributed to pattern matching 
were substantially low.

\subsection{RQ1: How often are warning suppression features used?}
\label{subsec:rq1}

By examining 1\,425 repositories that utilize 
only Findbugs~(1\,327), Spotbugs~(176), or both tools~(78),
we found that 168~(13\%) of them contain 
at least one filter file warning suppression,
with 132 projects being based purely on Findbugs,
12 purely on Spotbugs, 
and 22 on both.
Similarly, 91~(7\%) of the analyzed projects 
used at least one warning suppressing annotation,
with 45 of them being only Findbugs-dependent,
9 being only Spotbugs-dependent
and 28 dependent on both.
In total, 20~(2\%) of these projects used both
warning suppression features,
signifying that the tools examined but also their 
warning suppression features
are not mutually exclusive.
Furthermore, Findbugs seems to be the most popular tool,
while Spotbugs is only used in 12\% of the projects,
and often co-exists with Findbugs,
showing that the transition 
from Findbugs to Spotbugs
is not yet complete.


\begin{figure}[t]
  \includegraphics[width=\linewidth]{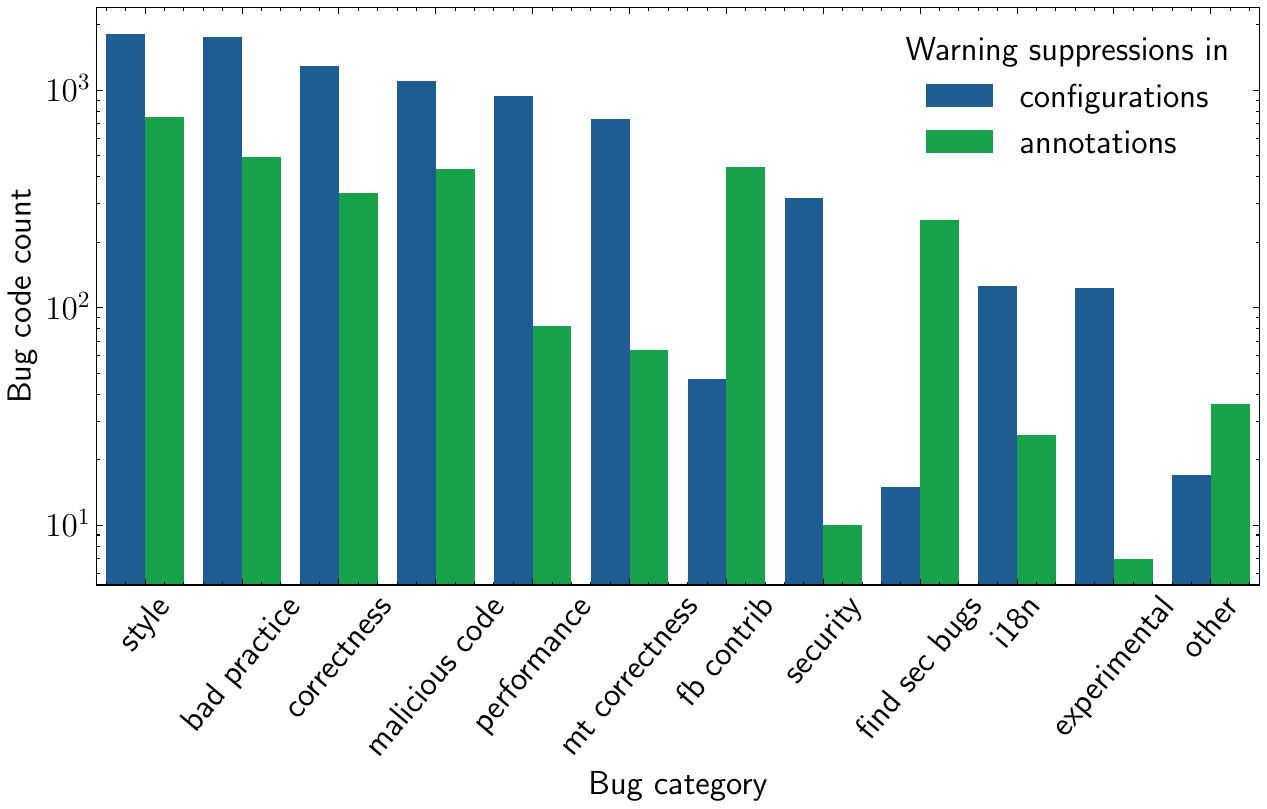}
  \caption{Warning suppressions per bug category found 
  inside warning suppression configurations and annotations}
  \label{fig:suppressions_per_category}
\end{figure}

A further analysis 
of the popularity of warning suppressions,
seen in Figure~\ref{fig:suppressions_per_category}, 
reveals the patterns developers tend to follow. 
Warnings pertaining to 
\emph{bad practices} and \emph{style} 
are most often suppressed in warning suppression configurations,
accounting for 22\% each. 
Significantly fewer suppressions are linked to \emph{security} 
(4\%), \emph{internationalization} (2\%), \emph{experimental} (1\%)
and \emph{plugin-related} warnings ($<$1\%).
On the other hand, in source code annotations, 
\emph{style}-related warnings are most commonly suppressed, 
followed closely by \emph{plugin-related warnings}.
\emph{bad practices} and \emph{malicious code} follow, 
with 17\% and 15\% respectively,

Most categories follow a similar pattern
in terms of their frequency of suppression
across the two suppression methods,
with the exception of
\emph{performance}, 
\emph{multi-threaded correctness}, 
and \emph{plugin-related} warnings.
In the first two cases, suppression through annotations 
is significantly less likely to occur
compared to their occurrence via configurations.
In the case of \emph{plugin-related} warnings,
the opposite is true.

\begin{figure*}[t]
  \includegraphics[width=\linewidth]{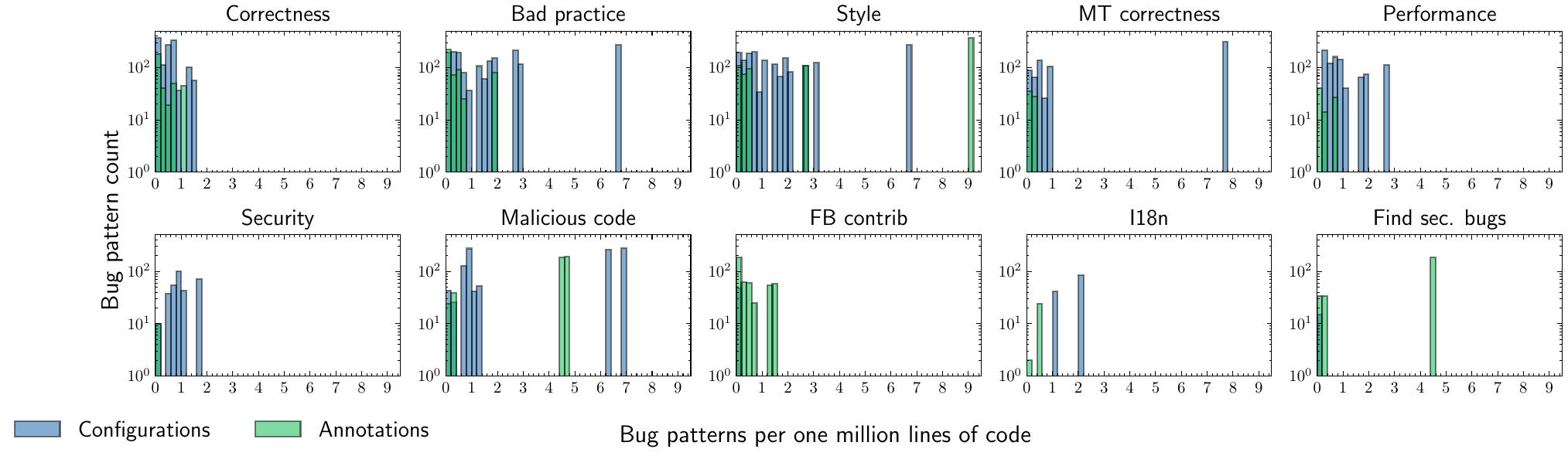}
  \caption{Distribution of total bug pattern warning suppressions per one million lines of code per bug category for warning suppressing configurations and annotations.}
  \label{fig:refs_per_code_annotations}
\end{figure*}

The number of bug pattern warning suppressions 
per one million lines of code 
for each bug category is also calculated, 
seen in Figure~\ref{fig:refs_per_code_annotations}. 
Most categories follow a similar distribution 
in both filter file and annotation induced suppressions, 
being skewed towards the lower values,
centered around zero and three suppressions 
per million lines of code.
Furthermore, filter file suppressions tend to be 
greater in terms of both mean references
and total observations.
Exceptions to the rules above are the \emph{malicious code}, 
\emph{style}, and \emph{plugin-related} categories,
where the mean value of warning suppressing annotations is higher than that of configurations.
This however, involves only a single or a few bug codes referenced repeatedly.
Similar outliers of frequent references are observed in the \emph{bad practice},
\emph{multi-threaded correctness}, \emph{malicious code} categories,
of filter file suppressions.

\begin{table}[!ht]
  \centering
  \caption{The scope of warning suppressions.}
  \begin{tabular}{lrrrrrr}
  \hline
  \toprule
      ~ & \multicolumn{2}{c}{\textbf{Config. Files}} & \multicolumn{2}{c}{\textbf{Annotations}} & \multicolumn{2}{c}{\textbf{Total}}\\
      ~ & count & \% & count & \% & count & \% \\
    \midrule
      Project & 1\,002	& 11  & ~	      & ~   & 1002	  & 8   \\
      Package & 849     & 9   & 0      & $<$1& 849     & 3 \\
      Class   & 3\,391  & 41  & 1\,152  & 39  & 4\,543  & 40  \\
      Method  & 2\,873  & 35  & 1\,405  & 48  & 4\,295  & 38  \\
      Field   & 690     & 8   & 336     & 11  & 1026    & 9   \\
      Other   & 0       & $<$1  & 50      & $<$1  & 50      & $<$1  \\
      \midrule
      \textbf{Total}   & 8\,298  & ~   & 2\,943    & ~   & 11\,240 & ~   \\
      \bottomrule
  \end{tabular}
  \label{tab:suppressions_per_scope}
\end{table}

The scope of warning suppressions is also examined 
in Table~\ref{tab:suppressions_per_scope}, 
showing that the majority of exclusions 
occur at \emph{class}~(41\%) scope 
in the context of warning suppressing configurations, 
and \emph{method}~(48\%) scope in the case of annotations. 
The results also show that annotations tend to be applied 
more judiciously and have a more limited scope than configurations.

\subsection{RQ2: Which warning suppression features are used in static analysis?}
\label{subsec:rq2}
The analysis of the 8\,298 configuration
and 2\,943 annotation bug patterns
revealed the existence of 
512 and
386 distinct bug codes respectively.
These codes fell into 12 separate categories, 
inclusive of the plugins \emph{find security bugs} 
and \emph{FB contrib.}, 
depicted in Figure~\ref{fig:suppressions_per_category}.

In relation to the plugins, 
\emph{find security bugs} 
\emph{FB contrib.} were used 
in 6~(3\%) 
and 5~(2\%)
out of 209 projects utilizing warning suppression configurations respectively.
Similarly, in the 95 projects utilizing annotations,
these plugins were observed in 6~(6\%) projects each.

\begin{table}[htbp]
  \centering
  \caption{Warning suppression categories for the first quantile of most referenced bug codes, for warning suppressing configurations and annotations.}
  \begin{tabularx}{0.47\textwidth}{lrrrrrr}
    \toprule
    \textbf{Category} & \multicolumn{2}{c}{\textbf{Config. Files}} & \multicolumn{2}{c}{\textbf{Annotations}} & \multicolumn{2}{c}{\textbf{Total}} \\
    & count & \% & count & \% & count & \% \\
    \midrule
    Style & 25 & 20 & 17 & 18 & 42 & 19 \\
    Bad practice & 23 & 18 & 19 & 20 & 42 & 19 \\
    Correctness & 28 & 22 & 13 & 14 & 41 & 18 \\
    Malicious Code & 16 & 13 & 7 & 7 & 23 & 10 \\
    Performance & 17 & 13 & 5 & 5 & 22 & 10 \\
    FB contrib. & 0 & $<1$ & 21 & 22 & 21 & 9 \\
    MT correctness & 8 & 6 & 4 & 4 & 12 & 5 \\
    Security & 8 & 6 & 0 & $<1$ & 8 & 4 \\
    Find sec. bugs & 0 & $<1$ & 6 & 6 & 6 & 3 \\
    I18N & 2 & 2 & 1 & 1 & 3 & 1 \\
    Other & 0 & 0 & 3 & 3 & 3 & 1 \\
    Experimental & 1 & 1 & 0 & $<1$ & 1 & $<1$ \\
    \midrule
    \textbf{Total} & 128 & & 96 & & 224 & \\
    \bottomrule
  \end{tabularx}
  \label{tab:q1_categories}
\end{table}

Table~\ref{tab:q1_categories} displays the distribution 
of warning suppression feature categories for the first percentile 
of most referenced bug codes, for warning suppression configurations and warning suppression annotations.
In terms of frequency, the categories of 
\emph{style}, 
\emph{bad practice}, 
and \emph{correctness} rank highest, 
representing 19\%, 19\%, and 18\% of total suppressions, 
respectively. 
This indicates that these types of warnings, 
potentially due to their broad, context dependent, and sometimes subjective nature, 
are most often deemed unactionable or difficult to fix.

The \emph{malicious code} category also stands out, 
especially in warning suppressing configurations, 
where it accounts for 13\% of the total observations, 
suggesting that warnings related to potential vulnerabilities 
are often suppressed in this method.

Interestingly, bug codes from the \emph{FB contrib.} plugin 
are predominantly suppressed via annotations, 
amounting to 22\% of total annotation suppressions. 
Conversely, there are almost no corresponding warning suppression configurations. 
This implies that warnings introduced by plugins 
are more prone to suppression via annotations.

The \emph{security} category, 
is relatively less suppressed, 
especially via annotations. 
This is a promising finding, 
suggesting that developers 
take security-related warnings more seriously.

In Figure~\ref{fig:refs_per_code_annotations} 
summarized in the previous section, 
several outliers become apparent when examining 
bug codes encountered across warning suppressing features.
The most frequently detected bug in both datasets is   
\emph{URF: unread public/protected field},
signifying similar priorities between the two warning suppression features. 
The recurrent appearance of both 
\emph{EI: expose rep.} and 
\emph{EI: expose rep. 2} in both 
warning suppression configurations and features 
suggests that warnings related 
to the exposure of the code's internal representation 
are often perceived as either unactionable 
or overly complex to address.

Notably, bug codes introduced by the \emph{FB contrib.} plugin, 
specifically \emph{MDM: thread yield} 
and \emph{PRMC: likely redundant method call}, 
are uniquely prevalent in the annotations. 
This implies that warnings introduced by plugins 
are often addressed via annotation-based suppressions. 
These bug codes are commonly associated with malpractices 
like misuse of \ttt{thread.yield()}, \ttt{thread.sleep()} methods, 
and redundant method calls, 
which are regularly suppressed.

\subsection{RQ3: For what purposes are static analysis warning suppression annotations used? \& RQ4: How could the use of static analysis warning suppression annotations be avoided?}
\label{subsec:rq3}
\begin{table}[!ht]
  \centering
  \caption{Category frequencies derived from manual coding of warning suppressing feature samples.}
  \begin{tabular}{lrrrr}
  \hline
  \toprule
      \textbf{Category} & \multicolumn{2}{c}{\textbf{Config. Files}} & \multicolumn{2}{c}{\textbf{Annotations}} \\
      ~ & count & \% & count & \% \\
  \midrule
    \textbf{False Positive}                 & 17	  & 5   & 15  & 4 \\
    \textbf{Technical Debt}                 & 101   & 27  & 101  & 34 \\
      \hspace{1em}Short-term Technical Debt & 53    & 14   & 96  & 28 \\
      \hspace{1em}Long-term Technical Debt  & 48    & 13  & 20  & 6  \\
    \textbf{Unactionable}                   & 250   & 68  & 209 & 61 \\
      \hspace{1em}Wrong Tool Assumption     & 148   & 40  & 115 & 34 \\
      \hspace{1em}Testing                   & 25    & 7  & 55  & 16 \\
      \hspace{1em}Outside of Analysis Scope & 77    & 21  & 39  & 11 \\
  \midrule
      \textbf{Total}                        & 368  & ~  & 340    & ~ \\
    \bottomrule
  \end{tabular}
  \label{tab:suppression_manual_coding}
\end{table}

Through the categorization described 
in Section~\ref{subsec:sampling-and-open-coding}, 
370 exclude filter matches 
and 340 annotation matches were examined. 
Our findings regarding 
the uses of static analysis warning suppression annotations 
are summarized in Table \ref{tab:suppression_manual_coding}.

Suppressions were assigned one category and one subcategory,
based on the rationale behind their usage (Figure~\ref{fig:category_structure}).

It is worth noting that the two suppression methods had a similar distribution
among the bug categories. 
\emph{Unactionable warnings} had the largest count,
followed by \emph{technical debt} and \emph{unactionable}.
\emph{Technical debt} was more common in annotations.
Regarding subcategories, 
\emph{short-term technical debt} was more common in both methods,
but was encountered almost as often as 
\emph{long-term technical debt} in warning suppressing configurations.
Configurations also seemed to ignore warnings \emph{outside of analysis scope}
more often.
This is justified by the fact that
warning suppressing annotations are more precise
while configurations can be used to suppress 
a larger set of warnings
with just a single rule.

\section{Discussion and Implications}
\label{sec:discussion_implications}

The use of warning suppression features
in static analysis tools 
points towards the existence of broader issues 
such as tool inaccuracies, 
improper configurations, 
and misapplications.
Common suppression patterns like \emph{testing} 
or application on~\emph{irrelevant contexts}, 
further imply a need 
for better tool settings 
to filter out unrelated warnings.
In our study, 
we note a lower rate of false positives~(5\%) 
than previous work~\cite{kremenek2003z,shen2011efindbugs,lenarduzzi2021critical}. 
Interestingly, we observed that 
silenced warnings are less likely to refer to false positives. 
Consequently, our findings suggest 
the need for resources to be directed towards 
improving report quality, 
automating error correction, 
and providing better support for developers, 
rather than solely focusing on tool precision.
This will encourage the timely resolution of warnings, 
and minimize dependency on warning suppression


The majority of exclusions 
occur at the \emph{class} scope~(41\%) 
in the context of warning suppression configurations, 
and at the \emph{method} scope~(48\%) in the case of warning suppression annotations. 
This finding suggests that most produced warnings 
are in some way useful 
and are not broadly dismissed 
at a project level.

Contrary to our initial expectations, 
the majority of filter file suppressions 
did not target the entire project, 
suggesting that developers 
are making efforts to address warnings 
at a more granular level. 
This observation highlights
the potential for static analysis tools 
to be more useful and less intrusive 
if their warning generation could be refined.

The disparity that exists between the coverage scopes 
of these two suppression methods
can be attributed 
to their distinct nature: 
annotations, due to their direct application, 
offer a more swift and straightforward solution 
to select bug patterns,
whereas configurations facilitate more comprehensive rules 
capable of suppressing 
a wide range of bug patterns simultaneously, 
potentially even on a global scale. 
Therefore, it is more likely to encounter a filter file rule 
that excludes a group of bug patterns rather than an annotation.

However, the situation changes when considering 
the coverage of tool plugins. 
In this context, 
the coverage of \emph{find security bugs}
and \emph{FB contrib.} appears substantially reduced.
This outcome is anticipated, 
given that these tools almost double 
the number of relevant bug patterns 
collectively while being utilized 
by only a fraction of the studied projects. 
Moreover, it is more likely to find
annotations rather than configurations
used to suppress plugin warnings,
which might be explained 
by the precision that annotations provide. 
As plugins offer a specialized set of checks 
typically related to specific sections 
of a project's code, 
a more explicit mechanism is preferable 
for managing these warnings.



With respect to the frequency of 
bug pattern suppression features 
within a project, 
the study finds that the majority
occur no more than three times 
per million source lines of code.
Conversely, a smaller set of bug patterns
are referenced multiple times
exhibiting a fat tail distribution.
This repetitive suppression 
of particular warning fixes 
can be attributed to several possible factors. 
These could range from 
the existence of \emph{false positives} 
due to inaccuracies in the tool's bug pattern detection, 
to developers deliberately 
allowing certain bug patterns to persist, 
potentially contributing to the accumulation 
of \emph{technical debt}. 
Importantly, the existence of 
such sporadic high-frequency outliers 
suggests a need for tool maintainers 
to concentrate their improvement efforts 
on a select number of bug patterns 
that are more frequently suppressed or ignored.
This leads us to the first implication.

\begin{implication}
  \label{implication:few_too_many}
  A large number of bug pattern warning suppressions
  are rarely referenced 
  while a small number of bug patterns 
  receive the majority of suppressions,
  suggesting that tool maintainers 
  should focus on improving the accuracy 
  and relevancy of these frequently suppressed
  bug patterns first.
\end{implication}
\smallskip

This implication means that tool maintainers 
should revise bug detection algorithms, 
focusing on the categories 
that experience high suppression rates. 
Moreover, they should provide clear guidance to developers 
regarding the optimal and relevant scopes of tool application, 
especially for categories that experience low suppression rates, 
to maximize the tool's utility and efficiency in improving code quality.


The category distribution data 
outlined in Sections~\ref{subsec:rq1} and~\ref{subsec:rq2} 
reveal several interesting aspects of 
how developers use static analysis tools 
to suppress warnings. 
With categories such as 
\emph{style}, 
\emph{bad practice}, 
and \emph{correctness} dominating warning suppressions, 
this could suggest that the tools 
currently being used generate 
a large number of false positives in these areas. 
Consequently, developers may perceive warnings in these categories 
as less accurate or less applicable, 
resulting in a higher number of suppressions. 
It also indicates the subjective nature of these categories, 
which may not align with a team's coding standards or the specific context of the project.

The distinctive behavior observed for the~\emph{FB contrib.} plugin, 
where warnings are primarily suppressed through annotations, 
could indicate a discrepancy between the issues 
the plugin identifies and the developers' priorities. 
This could also suggest that 
the warnings produced by the plugin 
might be more nuanced or less understood, 
prompting developers to suppress them more frequently.

Additionally, the difference in the treatment 
of \emph{security} and \emph{malicious code} categories 
is also intriguing. 
The apparent attention to the \emph{security} category hints 
towards a general awareness and sensitivity 
regarding security issues among developers. 
On the other hand, the more frequent suppression of 
\emph{malicious code} warnings, especially in 
warning suppressing configurations, 
raises questions about whether these warnings 
are considered less crucial or less actionable.

Regarding the reasons behind the use of 
tool warning suppression annotations
most of them were categorized as \emph{unactionable}.
This primarily involves \emph{wrong tool assumptions}, 
where the tool accurately identifies a potential code smell pattern, 
but some or all of its assumptions 
do not hold true for the specific case, 
or the suggested fix would be over-engineered 
and not justify the effort required.

A prime example of this is the 
\emph{HE: equals no hashcode} bug pattern. 
While the tool accurately warns 
when a class overrides equals(Object), 
but does not override hashCode(), 
this warning can be irrelevant 
if the equals method is implemented via a different mechanism 
than hashcode comparison. 
Instances were also found where 
the warning-triggering code was correct, 
but the proposed fix could make the code more complex, 
without offering substantial benefits. 
This suggests that many static analysis tools 
rely on overly restrictive conditions 
that may not improve overall code quality 
in the eyes of developers.

To mitigate the occurrence of such unactionable warnings, 
language constructs like \@NotNull and \@Nullable annotations 
could be used to guide static analysis, 
improving the accuracy of the pattern matchers. 
The sensitivity of the analysis could be adjusted 
to incorporate bug patterns 
with higher severity 
at the cost of analysis recall. 
Furthermore, bug descriptions could be split 
into multiple subcategories, 
each with its own pattern matcher, 
to enhance specificity and accuracy. 
The descriptions could also include example fixes 
to serve as a guideline for developers, 
a feature currently absent in most bug descriptions.
These insights lead to the following implication.

\begin{implication}
  \label{implication:unactionable}
  The prevalence of \emph{wrong tool assumption} warnings 
  suggests the need for tool refinement to improve 
  effectiveness, accuracy, and documentation. 
  Mitigating these could involve more precise 
  bug code definitions, 
  optimized tool configurations, 
  improved code annotation, 
  and programming language enhancements.
  \end{implication}
  \smallskip

Similarly, many warnings were categorized as \emph{technical debt}. 
These types of warnings were more frequently found 
in warning suppression annotations than in configurations. 
Moreover, annotations appear to focus more on 
\emph{short-term technical debt}, 
while suppression configurations on \emph{long-term technical debt}. 
This difference can be attributed to the nature of each warning: 
annotations often quickly mask warnings whose fix may be simple, 
but not worth the effort to address immediately. 
They might also serve as reminders for developers 
to address the issue in future iterations.

Warning suppression configurations, on the other hand, 
apply to a broader set of warnings 
and code fragments, 
masking changes that are likely more complex 
and require significant refactoring. 
A direct approach to reducing the number of warnings 
attributed to technical debt 
would involve programmers spending more time 
resolving simpler issues. 
The static analysis tool 
could also provide a more detailed description 
of the warning causes and recommended fixes 
to encourage developers to address the problem 
when the warning is generated, 
rather than delaying its resolution. 

As a recommendation, 
annotations should only be used for \emph{short-term technical debt} 
and should be resolved as early as possible, 
while \emph{long-term technical debt} 
should be managed at the project level 
instead of being annotated. 
These findings lead to the following implication.

\begin{implication}
  \label{implication:technical_debt}
  The abundance of technical debt could 
  indicate low code quality and procrastination, 
  but also a lack of guidance from the tool. 
  Mitigating this would require improving code quality standards 
  to the extent possible, 
  while making static analysis warnings more descriptive 
  and providing fix recommendations 
  to prompt more immediate resolution.
  Tools that automate the fixes 
  could also be employed or developed.
\end{implication}
\smallskip

Additional common instances of warning suppression were identified in 
\emph{testing} and 
\emph{outside of analysis scope} scenarios. 
The former was more prevalent in warning suppressing annotations, 
while the latter was common in configurations. 
Such suppressions are often used for warnings 
found in test cases 
or in irrelevant contexts, 
such as non-Java files 
and external libraries. 
This suggests that the tool was sometimes 
applied in the wrong context, 
as many of the code quality requirements 
do not apply to test code or file types 
of different languages or formats.

Interestingly, annotations
were commonly found in \emph{testing} cases. 
This could suggest that developers 
may not realize that many static analysis 
assumptions do not hold for test code. 
Once the warning occurred, 
the developer might recognize that the it is 
\emph{unactionable}, 
suppress it, 
and not take the time 
to properly reconfigure 
the analysis to ignore test code. 
Consequently, instead of suppressing warnings 
that may occur in these cases through annotations, 
the tool should be configured 
to selectively ignore test code and other irrelevant scenarios. 
It might also be worth improving test frameworks 
to minimize the need for cases that trigger such warnings. 
This practice could not only decrease warning irrelevance 
but also improve the overall performance of the analysis, 
leading to the following implication.

\begin{implication}
\label{implication:testing}
Test cases 
and utility source files (i.e. automatically generated files) written in other languages, 
though part of the project's code, 
often do not adhere to the same code quality requirements. 
Static analysis should be properly configured 
through the tool's configuration files, 
instead of annotations, 
leading to cleaner code and improved analysis performance.
\end{implication}
\smallskip

In order to reduce the use of static analysis warning suppression features, 
it is imperative to possess a comprehensive understanding 
of the categories of warnings and the rationale behind each. 
This entails mitigating the number of false positives, 
promoting a codebase environment 
that encourages the timely resolution 
of identified warnings, 
and enhancing the specificity 
and scope of our tooling. 
These collective efforts 
can significantly decrease 
the reliance on warning suppression configurations 
and annotations .

Remarkably, the rate of false-positive warnings 
in our study was considerably lower (13\%) 
than the lowest estimate in previous work~\cite{kremenek2003z,shen2011efindbugs,lenarduzzi2021critical}. 
This discrepancy could be attributed to our focus 
on warning suppressions 
rather than all warnings 
generated by the tools. 
To alleviate the effect of false positives, 
improvements in the quality of data flow analysis 
and pattern matchers are necessary. 
However, this might slow the analysis down, 
making it impractical for larger projects.
This leads us to Implication~\ref{implication:true-false-positives}.

\begin{implication}
\label{implication:true-false-positives}
Warnings that are ignored 
are significantly less likely 
to refer to static analysis defects (false positives) 
compared to warnings that are difficult to be acted upon. 
Hence, efforts to improve tools 
would be better invested 
in improving reporting, 
automated error correction, 
and guidance to software practitioners, 
rather than increasing tool accuracy.
\end{implication}
\smallskip


Finally, the results of our study demonstrate 
a heavy reliance on warning suppression features, 
indicating a need for alternative solutions to manage warnings. 
This overuse of suppressions 
could be attributed to either 
the lack of understanding about the warnings themselves 
or the inadequacy of the tools 
to adequately communicate the nature 
and impact of the warnings. 
Relevant literature has proposed 
assigning the responsibility of resolving warnings
to a dedicated developer team as a solution, 
to promote a uniform way of addressing unactionable warnings, 
but has not been widely adopted~\cite{ayewah2007evaluating}.
Similar recommendations include temporary warning suppression,
to allow developers to focus on more pressing issues~\cite{6606613}.
Improving the tool's provided information towards the developer
through IDE integrations~\cite{10.1145/1879211.1879216}
like heatmaps~\cite{168856} or text popups
is another approach for tackling this issue.

Throughout the categorization process, 
we observed a low quality in the descriptions
of bug patterns.
Improvements in the warning messages, 
inclusion of additional context for warnings,
or even integration of interactive features
as done in similar static analyzers,
like ESLint,\footnote{\url{https://eslint.org/docs/rules/}}
would facilitate a deeper understanding of the warnings.
This enhanced communication would 
not only reduce the dependency 
on warning suppressions, 
but also potentially lead to a cleaner, 
safer codebase.

\begin{implication}
  \label{implication:warning-education}
  Improving the communicative and educational aspects 
  of static analysis tools 
  can significantly reduce the need for warning suppression annotations. 
  As developers gain a better understanding of the warnings, 
  they are more likely to address them directly
  or properly configure their tools in advance, 
  instead of resorting to suppressions.
  \end{implication}

\section{Threats to Validity}
\label{sec:threats}

The study's external validity as far as generalizability is concerned
suffers by examining only two widely used static analysis tools, Findbugs and Spotbugs,
targeting a single language, Java,
rather than a larger array of similar software.
Despite this, when evaluated in comparative studies,
both analyzers are able to capture an equally wide range of defects
as similar tools,
while many related studies have also deemed them representative.
Furthermore, results only target Java-based projects,
and therefore some categorization labels do not always apply to 
tools that target other programming languages.

There are also concerns regarding the internal validity 
of the study due to the limited number of samples analyzed. 
The canonicalization of warning suppressions, 
in combination with the sampling technique used, 
resulted in homogeneous observations,
but may have compromised the accuracy of the context. 
For example, 
the reasoning behind suppressing an entire bug category 
may differ from that of suppressing 
a specific bug pattern within the same category.

Moreover, external validity could suffer from the steps of the study's methodology
that involve subjective judgment. 
These include the classification of true false-positive annotations,
the labeling of the reasons why static analysis warning-suppressing
annotations were used, 
and the resolution of the way to avoid the use of each examined annotation.
To mitigate this, the authors have 
verified the results of the manual coding process
through Cohen's Kappa inter-rater agreement coefficient,
which was found to be substantial for
both filter file ($\kappa = 0.811$)
and annotation ($\kappa = 0.895$) categorizations.

Another potential limitation 
relates to the use of regular expressions 
to identify and analyze 
bug pattern suppression annotations 
in the projects' source code. 
While using advanced methods of program analysis
(i.e. using a program's abstract syntax tree) 
could offer better parsing accuracy, 
it would likely slow down 
the implementation and analysis process
to infeasible levels. 
However, it's worth noting that 
only a handful of results 
were omitted from the final dataset
due to parsing errors.

\section{Conclusions and Future Work}

This study extends prior work by carrying out 
an empirical evaluation 
of popular Java-based static analysis tools,
analyzing the reasons
static analysis warning-suppressing annotations 
and configurations are employed.
The analysis indicates that almost all warnings 
could be suppressed,
but it's typically only a small portion of 
bug patterns that get suppressed in each project.
This pattern is more prominent 
when examining source code annotations, 
whereas configurations usually address a larger group of warnings.

Contrary to the findings of prior studies, 
false positives 
only made up a minor proportion 
of the analyzed suppressions. 
However, there was a substantial number of suppressions 
that introduced technical debt, 
which could suggest a disregard for code quality 
on the part of developers 
or a lack of guidance from the tool. 
There were also numerous suppressions 
caused by misleading suggestions and incorrect assumptions, 
indicating a need for improved bug pattern definition 
by tool maintainers and better code annotation from the users.
Lastly, instances were discovered 
where test cases resulted in unactionable warnings
from Findbugs and Spotbugs, 
emphasizing the need for software practitioners 
to properly configure static analysis in their projects.

Future research in this area can examine 
how these patterns of suppression 
vary across different static analysis tools and programming languages, 
to investigate whether the usage patterns 
identified in this study 
are universal 
or specific to Findbugs and Spotbugs in Java. 
Moreover, a similar approach could be taken to explore 
the use of suppression annotations in dynamically typed languages. 
A deeper understanding of the decision-making process 
of developers when deciding to suppress warnings 
would also be valuable, 
possibly through surveys 
or interviews with software practitioners 
to gather qualitative data. 
Finally, in line with the findings of this study, 
future work could explore more sophisticated approaches 
to the automatic detection and categorization 
of suppression annotations in the source code 
to further refine the accuracy of these investigations.

\section{Data Availability}
\label{sec:availability}
The anonymized dataset,
that supports the findings of this study 
is available on Zenodo\footnote{10.5281/zenodo.10119397}.

\bibliographystyle{plainnat}
\bibliography{lps2023.bib}

\end{document}